# Gravity-capillary waves on the free surface of a liquid dielectric in a tangential electric field

**Evgeny A. Kochurin**

Institute of Electrophysics, Ural Branch of Russian Academy of Sciences
620016, 106 Amundsen str., Yekaterinburg, Russia

and **Nikolay M. Zubarev**

Institute of Electrophysics, Ural Branch of Russian Academy of Sciences
620016, 106 Amundsen str., Yekaterinburg, Russia
P.N. Lebedev Physical Institute, Russian Academy of Sciences
119991, 53 Leninskij prospect, Moscow, Russia

## ABSTRACT

Processes of propagation and interaction of nonlinear gravity-capillary waves on the free surface of a deep non-conducting liquid with high dielectric constant under the action of a tangential electric field are numerically simulated. The computational method is based on the time-dependent conformal transformation of the region occupied by the fluid into a half-plane. In the limit of a strong electric field, when the gravitational and capillary forces are negligibly small, there exists an exact analytical solution of the electro-hydrodynamic equations describing propagation without distortions of nonlinear surface waves along (or against) the electric field direction. In the situation where gravity and capillarity are taken into account, the results of numerical simulations indeed show that, for large external field, the waves traveling in a given direction tend to preserve their shape. In the limit of a strong electric field, the interaction of counter-propagating waves leads to the formation of regions, where the electrostatic and dynamic pressures undergo a discontinuity, and the curvature of the surface increases infinitely. The Fourier spectrum of the surface perturbations tends to the power-law distribution $(k^{-2})$. In the case of a finite electric field, the wave interaction results in a radiation of massive cascade of small-scale capillary waves that causes the chaotic behavior of the system. The investigated mechanism of interaction between oppositely-traveling waves can enhance development of the capillary turbulence of the fluid surface.

## 1 INTRODUCTION

It is well known [1] that an external electric field directed tangentially to the unperturbed free or contact surfaces of dielectric liquids has a stabilizing effect on the boundary. On the contrary, the normal field results in an aperiodic growth of the boundary perturbations [2-4]. In recent years, new experimental results concerning the influence of electric field on dynamics of capillary waves on the surface of dielectric liquids have been obtained [5,6]. The applied interest in studying the liquid surface dynamics in external electric field is related to the possibility of controlling the behavior of liquid surfaces and suppressing hydrodynamic instabilities [7-10]. The features of the nonlinear evolution of capillary waves at the liquid interfaces in the presence of the horizontal field were analyzed in [11-13]. It has been shown in [14-16] that, in the case of a strong electric field (the effects of gravitational and capillary forces are negligibly small), nonlinear waves on the surface of a liquid with high dielectric constant can propagate without distortions along (or against) the field direction.

As a rule, the description of nonlinear traveling waves preserving their form (the so-called progressive waves) imposes significant restrictions on the shape of perturbations. Solitary waves [17] and periodic Crapper's waves [18] are the examples of such nonlinear perturbations. In this connection, the solutions obtained in [14-16] are not typical: they describe propagation of waves of an arbitrary configuration. The main physical restriction for their applicability is related to the value of the external electric field: it must be large enough to neglect the effects of gravity and capillarity. In the present work, processes of propagation and interaction of nonlinear surface waves in the framework of the full system of hydrodynamic equations will be numerically investigated. In particular, it will be shown that the limit of a strong field [14-16] can be realized.

For the description of nonlinear dynamics of the fluid surface, we will use the method of time-dependent conformal transform (the region occupied by the liquid is mapped into a half-plane). This approach was developed in [19,20] in studying nonlinear waves on the surface of liquids in the absence of an external electric field. The method was used in [13,21,22] for the study of electro-hydrodynamics of liquid dielectrics with free surfaces. At present time, the computational techniques based on the conformal transformations develop intensively; see, for example, [23, 24]. The main advantage of this approach in comparison with the classical finite-difference schemes is in reducing the number of spatial variables. Together with the use of spectral methods for calculating the spatial derivatives, the approach allows conducting direct numerical simulation of electro-hydrodynamics of liquids with high efficiency and accuracy.

## 2 LINEAR ANALYSIS OF THE PROBLEM

The dispersion relation for linear waves on the surface of a perfect dielectric liquid (free charges are absent) under the action of the horizontal electric field has the following form [2]:

$$\omega^2 = gk + \frac{\varepsilon_0(\varepsilon-1)^2 E^2}{\rho(\varepsilon+1)}k^2 + \frac{\sigma}{\rho}k^3, \qquad (1)$$

where $\omega$ is the frequency, $k$ is the absolute value of wave-vector, $g$ is the acceleration of gravity, $E$ is the value of external electric field strength, $\varepsilon_0$ is the dielectric constant of vacuum, $\varepsilon$ and $\rho$ are the relative dielectric constant and, respectively, the mass density of the liquid, and $\sigma$ is the surface tension coefficient. Here and below in the paper, we consider plane symmetric waves propagating along the direction of the external electric field. It should be noted that this approximation is justified, since there is anisotropy in the problem related to the distinguished direction of the external field, see for more details [25].

In the absence of the external field, $E=0$, the relation (1) describes the propagation of plane gravity-capillary waves. It is easy to show that for the wavenumber $k_0=(g\rho/\sigma)^{1/2}$, the phase speed reaches the minimum value, $V_0=(4\sigma g/\rho)^{1/4}$, i.e., waves on the free surface of a liquid cannot propagate with the velocity less than $V_0$. As an example, the minimum phase speed of the gravity-capillary waves on the surface of water is $V_0 \approx 23$ cm/s; it corresponds to the wavelength $\lambda_0=2\pi/k_0 \approx 1.7$ cm. In the presence of the electric field, the phase speed $V_p$ of surface waves also has a minimum at the same wavelength $\lambda_0$. In this case, the minimum value of the speed depends on the electric field strength. It is easy to show that

$$V_p(k_0) = \left( 2\left(\frac{\sigma g}{\rho}\right)^{1/2} + \frac{\varepsilon_0 \varepsilon E^2}{\rho} \right)^{1/2}, \qquad (2)$$

where we take into account that, in the problem under study, $\varepsilon \gg 1$. The expression (2) shows that the external electric field leads to increase in the velocity of waves propagation.

The relation (2) allows to define the notion of a strong electric field in the problem. It is convenient to introduce the value of the electric field strength $E_0$, for which the electrostatic forces are comparable with the gravitational and capillary ones,

$$E_0^2 = \sqrt{\sigma g \rho} / \varepsilon_0 \varepsilon.$$

The limit of a strong field is realized in the case

$$\beta^2 = (E/E_0)^2 \gg 1,$$

where $\beta$ is a dimensionaless parameter defining the ratio of the external field to the specific value $E_0$. Let us discuss the feasibility of this limit. The estimations of $E_0$ for water and ethyl alcohol (these liquids have relatively high permittivities) give 1.9 kV/cm and 2.3 kV/cm, respectively. In the case of liquid - atmospheric gas interface the parameter $\beta^2$ is bounded above by Paschen`s law. It is known that the breakdown voltage in air under standard conditions is of the order of 33 kV/cm, so the parameter $\beta^2$ should not exceed about 200-300 dimensionaless units. As we will show further, the limit of a strong field can be realized with high accuracy for $\beta^2$=100, i.e., for the field less than the breakdown threshold.

In the conclusion of this section, let us discuss the important feature of the limit of a strong field. For $\beta \gg 1$, the terms responsible for the influence of gravity and capillarity can be neglected, and the relation (1) will describe the propagation of linear waves without dispersion. As it was noted above, in such a situation, exact solutions of the full electro-hydrodynamic equations have been found [14-16]. According to them, the nonlinear surface waves of an arbitrary shape can propagate without distortions along (or against) the direction of the external electric field. The interaction is possible only between counter-propagating waves (in [21], it was shown that such interaction is elastic). It should be noted that this situation is similar to that for the Alfven waves in an ideal conducting fluid. The wave packets of arbitrary forms can travel nondispersively with the Alfven speed in or against the direction of the external magnetic field [26] and their interaction is elastic too.

## 3 EQUATIONS OF MOTION

We consider a potential flow of an incompressible ideal dielectric liquid of infinite depth with a free surface in an external uniform horizontal electric field. The boundary of the liquid in the unperturbed state is the horizontal plane $y=0$ (the $x$ axis of the Cartesian coordinate system lies in this plane and the $y$ axis is perpendicular to it). Let the function $\eta(x,t)$ specify the deviation of the boundary from the plane; i.e., the equation $y=\eta$ determines the profile of the surface. Let the electric field be directed along the $x$ axis and $E$ in magnitude. As was shown in [14-16], the normal component of the electric field in the liquid, in the case of large permittivity, $\varepsilon \gg 1$, is much smaller than the tangential component. This means that field lines inside the liquid are directed along a tangent to its surface. In this case, the field distribution in the liquid can be determined disregarding the field distribution above it.

This situation is described by the following equations of motion. The velocity potential of the liquid $\phi$ and electric field potential $\varphi$ satisfy the Laplace equations

$$\nabla^2 \varphi = 0, \qquad \nabla^2 \phi = 0, \qquad y < \eta(x, y).$$

They should be solved together with the dynamic boundary condition (non-stationary Bernoulli equation)

$$\phi_t + (\nabla \phi)^2 / 2 = -(P_0 - P_E) / \rho - g\eta + \sigma K / \rho, \quad y = \eta, \quad (3)$$

where $P_E$ is the electrostatic pressure, $P_0$ is the energy density of the external electric field in the liquid, $K = \eta_{xx}/(1+\eta_x^2)^{3/2}$ is the curvature of the fluid surface. In the situation under study, the quantities $P_E$ and $P_0$ are defined by the following way: $P_E = \varepsilon_0 \varepsilon (\nabla \varphi)^2 / 2$ and $P_0 = \varepsilon_0 \varepsilon E^2 / 2$ (for more details, see [14]).

The dynamics of the surface is described by the kinematic boundary condition

$$\eta_t = \phi_y - \eta_x \phi_x, \qquad y = \eta. \qquad (4)$$

The potentials satisfy the condition $\varphi_y - \eta_x \varphi_x = 0$ at the boundary $y=\eta(x,t)$, and the conditions $\phi \to 0$, $\varphi \to -Ex$ at infinity $y \to -\infty$.

Together, the above relations are a closed equations system describing the motion of a deep dielectric liquid with free surface under the action of gravity, capillarity, and electrostatic forces caused by the external horizontal electric field. In the linear approximation, where the amplitude of surface perturbations is infinitesimal, this system reduces to the dispersion relation (1). It is convenient now to switch to dimensionless notations as follows:

$$y \to y/k_0, \ x \to x/k_0, \ t \to t\tau, \ \phi \to \phi/\tau k_0^2, \ E \to \beta E_0, \ \varphi \to \varphi \beta E_0/k_0,$$

where we introduced the quantity $\tau = \sqrt{2}/V_0 k_0$, determining the characteristic time scale of the system in the absence of electric field, $\tau \sim 0.01$ s.

## 4 EQUATIONS IN CONFORMAL VARIABLES

For complete description of the dynamics of the system, it is necessary to obtain distributions of the velocity field and electric field inside the liquid. Thus, the equations of motion given in the previous section have the dimension (2+1). It turns out that there is an effective way to reduce the dimension by transition to new spatial (conformal) variables. Such a transition allows to significantly increase the speed and accuracy of calculations.

Similar to [19-22], let us make the conformal transformation of the region occupied by the liquid into the half-plane: $\{x, y\} \to \{u, v\}$, where $v<0$. In terms of the complex variables $z=x+iy$ and $w=u+iv$, the conformality of the transformation implies that $z$ is an analytic function of $w$. The auxiliary variables $u$ and $v$ in the problem under study have clear physical meaning: $u$ coincides with the field potential $\varphi$ except for the sign and the condition $v$=const specifies the electric field lines. In the new variables, the Laplace equations for the electric field and velocity potentials can be solved analytically. As a result, the initial problem of motion of the liquid can be reduced to the problem of motion of its free surface, which has a lower dimension of (1+1).

Since the function $y(w)$ is an analytic, the surface of the liquid in the new variables is specified by the parametric expressions

$$y = Y(u,t), \qquad x = X(u,t) = u - \hat{H}Y,$$

where $\hat{H}$ is the Hilbert transform defined in Fourier-space as

$$\hat{H}_k = i \, \text{sign}(k).$$

Thus, the relation between the functions $\eta(x,t)$ and $Y(u, t)$ is given in the implicit form

$$Y(u,t) = \eta(u - \hat{H}Y, t).$$

The function $\Psi(u,t)$ specifying to the value of velocity potential at the boundary $v=0$ is introduced by the same way.

The procedure of obtaining the equations of motion in conformal variables is well known; see [19-22]. For this reason, we will not give a detailed derivation these equations, and just write them down in the final form. Let us introduce the complex functions

$$Z = X + iY, \qquad \Phi = \Psi + i\hat{H}\Psi,$$

which can be analytically continued into the lower complex half-plane. It is convenient to use the projection operator $\hat{P} = (1 + i\hat{H})/2$, which transforms a real function to an analytic complex function in the lower half-plane of the complex variable $u$, i.e., $Z = u + 2i\hat{P}Y$ and $\Phi = 2\hat{P}\Psi$. Let us now introduce the Dyachenko variables [19]

$$R = 1/Z_u, \qquad V = i\Phi_u/Z_u.$$

It should be noted that these functions can be interpreted in terms of the real physical variables. The function $|V(u,t)|$ corresponds to the absolute value of the fluid velocity at the boundary, and the quantity $\beta|R(u,t)|$ determines the local electric field strength on the surface in the case of $\beta \neq 0$. The function $|R(u,t)|^2$ has also a sense of the inverse Jacobian of the conformal transformation: $|R|^2 = J^{-1} = (X_u^2 + Y_u^2)^{-1}$.

As a result, the dynamic and kinematic boundary conditions (3) and (4) are rewritten as follows:

$$V_t = i(UV_u - D_u R) + (R-1) - 2Q^2\hat{P}(Q_u\overline{Q} - \overline{Q}_u Q)_u, \qquad (5)$$

$$R_t = i(UR_u - U_u R), \qquad (6)$$

where we introduced the notations

$$U = \hat{P}(V\overline{R} + \overline{V}R), \qquad D = \hat{P}(V\overline{V} - \beta^2 R\overline{R}), \qquad Q = R^{1/2}.$$

The overline stands for complex conjugation.

The equations (5) and (6) constitute the system of integro-differential equations describing the fully-nonlinear evolution of the free surface. It is completely equivalent to the original system of partial differential equations. For the numerical solution of equations (5) and (6), we will use the explicit Runge-Kutta method of the fourth order of accuracy with respect to time. All spatial derivatives and Hilbert operators will be computed by means of the fast Fourier transform, i.e., the boundary conditions will be periodic. For controlling calculation error, we can use the fact that the system under consideration is conservative, and its total energy has the following form:

$$H = -\frac{1}{2}\int_{-\infty}^{+\infty}(\Psi\hat{H}\Psi_u + \beta^2 Y\hat{H}Y_u - Y^2 X_u - 2[J^{-1/2} - X_u])du. \qquad (7)$$

Concluding this section, let us return to the strong field limit, where gravity and capillarity can be neglected. For the case $\beta \gg 1$, the system (5) and (6) takes the compact form

$$V_t = i(UV_u - D_u R), \qquad R_t = i(UR_u - U_u R). \qquad (8)$$

It is important that this system admits a pair of exact particular solutions, $V^{\pm}(u \mp \beta t) = \mp i\beta(R-1)$. These solutions correspond to waves of an arbitrary geometry that propagate at a constant velocity $\beta$ without distortions along (upper signs) or against (lower signs) the direction of the external electric field. In the next section, we will numerically investigate the possibility of realization of these solutions in the framework of the full system of equations (5) and (6). Also, we will consider the interaction between oppositely-propagating nonlinear waves.

# 5 SIMULATION RESULTS

## 5.1 TRAVELING WAVES

First of all, let us consider the possibility of propagation of stationary nonlinear waves on the liquid boundary in the framework of the complete equations (5) and (6). Since the gravity-capillary waves themselves are dispersive, let us set the initial conditions in such a way that a wave would be distorted only due to nonlinear effects (the linear dispersion is absent). The expression for the phase velocity in dimensionless variables has the form

$$V_p(k,\beta) = \omega/k = \sqrt{k^{-1} + \beta^2 + k}. \qquad (9)$$

Let us further consider the evolution of a periodic wave of the amplitude $A$ and the wavelength $2\pi/k_1$. In the linear approximation, its velocity is constant and equal to $V_p(k_1,\beta)$ according to (9). The corresponding initial conditions for the functions $R$ and $V$ are the following:

$$R(u,0) = 1 + A\exp(-ik_1 u), \quad V(u,0) = -AiV_p(k_1)\exp(-ik_1 u).$$

The deformation of this wave is intensified with an increase in the amplitude $A$.

The figure 1 shows the evolution of the liquid surface for the wave amplitude $A=0.5$ and different values of $\beta$. The vertical axis corresponds to time measured in the relative values $t/T_0$, where $T_0 = 2\pi/V_p(k_1,\beta)$ is the time period of the wave propagation. Since the wave speed depends on the field strength, it is convenient to compare the dynamics of the fluid boundary for various $\beta$ in the relative time scale. Figure 1 shows that electric field really has a stabilizing effect on the dynamics of a nonlinear perturbation of the surface. It can be seen that, for $\beta=0$, the deformation of the wave is very strong. At small field, $\beta^2=10$, the distortion of the wave occurs with less intensity, and for the relatively large electric field, $\beta^2=100$, the shape of the perturbation almost does not change during ten periods, $t=10T_0$. The obtained results indicate that the limit of a strong electric field can indeed be realized for the waves propagating in a single direction.

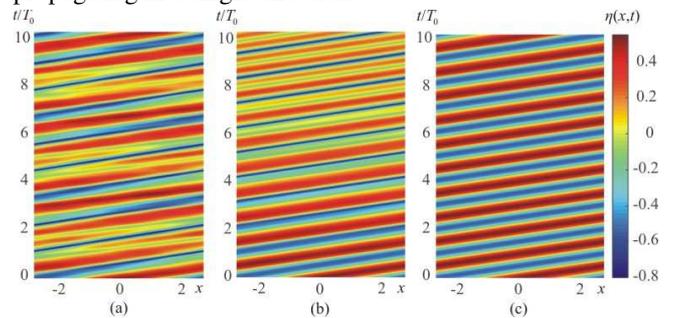

**Figure 1.** Evolution of the free surface for different values of the electric field strength: (a) $\beta = 0$, (b) $\beta^2 = 10$, (c) $\beta^2 = 100$.

It should be noted that the investigated situation is strongly nonlinear, since the amplitude of surface disturbance ($A=0.5$)

is close to the existence threshold for surface waves. For larger amplitudes of perturbations, the self-intersections of the boundary and the formation of air bubbles can arise at the liquid surface as shown in figure 2.

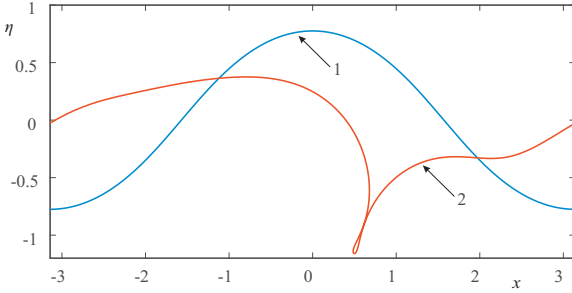

**Figure 2.** The calculated shape of the boundary is shown at the initial moment (blue curve "1") and at the end of calculation interval $t/T_0 \approx 0.76$ (red curve "2") for the parameters $A=0.65$, $k_1=1$, $\beta=0$.

## 5.2 INTERACTION OF WAVES IN THE STRONG FIELD LIMIT

As it was shown earlier, the limit of a strong field is realized for $\beta^2=100$. Although nonlinear waves individually propagating in the positive or negative direction of the $x$ axis behave as linear, this does not mean that oppositely propagating waves do not interact with each other. The question on the system dynamics in the case of wave collisions is of principal importance in the problem under investigation. First of all, we consider elastic interaction of the waves in the framework of reduced system (8). Let us set the initial conditions as follows:

$$R(u,0) = 1 + A_1 \exp(-ik_1u) + A_2 \exp(-ik_2u), \quad (10)$$

$$V(u,0) = -A_1 i\beta \exp(-ik_1u) + A_2 i\beta \exp(-ik_2u).$$

The results of numerical simulation are shown in figure 3; the parameters are chosen as $A_1=0.25$, $A_2=0.1$, $k_1=1$, $k_2=2$, $\beta^2=100$. The calculations were stopped at the moment when the relative error in computation of energy (7) reached the value $10^{-6}$. The number of used Fourier harmonics was equal to $2^{15}$, the spatial period of the problem was $2\pi$. The integration with respect to time was carried out with the adaptive step, which minimum value was of the order of $10^{-7}$.

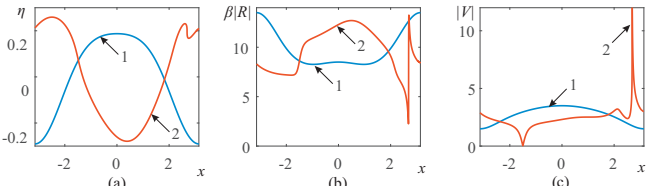

**Figure 3.** (a) The surface of liquid, (b) the local electric field, and (c) the velocity of the fluid surface are shown at the initial moment (blue curves "1") and at the end of calculation interval $t/T_0 \approx 4.61$ (red curves "2").

It can be seen that the region with steep wave front has been formed on the liquid boundary at the end of calculation interval, see figure 3a. In this region, the electric field pressure undergoes a discontinuity, figure 3b, and the dynamic pressure increases almost an order of magnitude, figure 3c. The spatial-temporal evolution of the electric field pressure is plotted in figure 4. As one can see, at the initial stages of the system evolution, the quantity $\beta|R|$ is smooth enough. At some moment of time, the interaction of surface waves leads to the formation of narrow spatial regions (shock fronts), where electric field changes sharply. It is interesting to note that the shock fronts do not move with constant velocity – see figure 4b. Let us explain the mechanism of this phenomenon.

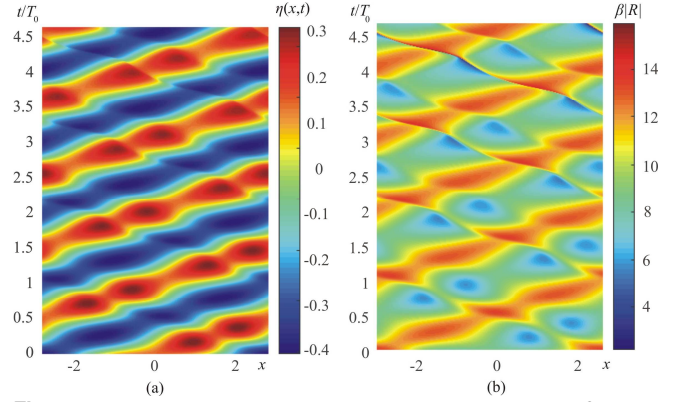

**Figure 4.** Evolution (a) of the free surface and (b) of the quantity $\beta|R|$.

In a situation where the shock fronts have been formed, the spatial scales of oppositely traveling waves differ significantly, see figure 4. For the wave with the amplitude $A_2$ and wavenumber $k_2$, the scale is defined by the width of the shock front $d$. For the other wave (with the parameters $A_1$ and $k_1$) the spatial scale equals the wavelength $2\pi/k_1$. Obviously, we have $2\pi/k_1 \gg d$. Such a situation was analyzed in [27], where it was shown that the velocity of a small-scale wave depends on geometry of the counter-propagating large-scale wave. From the physical point of view, this effect is related to the fact that the wave speed is proportional to the electric field strength in the limit $\beta \gg 1$. The shock front velocity is defined not by the external (unperturbed) field $\beta$, but by the local (perturbed) value of the field $\beta|R|$. The figure 4 demonstrates exactly this behavior: the velocity of the shock front is clearly correlated with the elevation $\eta$ and the local field $\beta|R|$. It should be noted that the velocity of the shock front does not depend on its shape and it is mostly defined by the geometry of large-scale wave, along which it propagates.

The spectrum of the function $Y(u,t)$ presented in figure 5a gives an evidence of the singular behavior of the system. It can be seen that during the system evolution the spectral functions of $Y(u,t)$ tend to a power-law distribution. Its exponent is close to two, which indicates to the formation of a singularity in the second spatial derivative of the function $Y$. From the physical point of view, such a behavior can correspond to an infinite increase in the curvature of the surface $K$. Figure 5b shows the time evolution of the maximum of absolute value of the surface curvature. It can be seen that the curvature increases jumpwise. Almost discrete jumps of the curvature are associated with the interaction of shock fronts moving in the opposite directions.

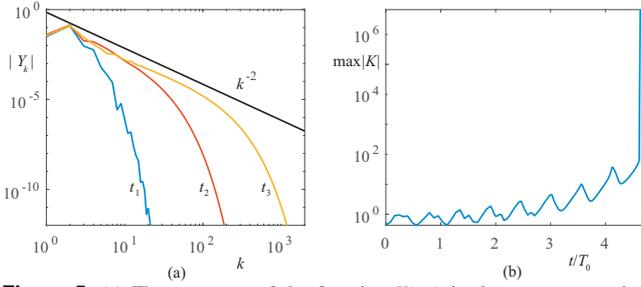

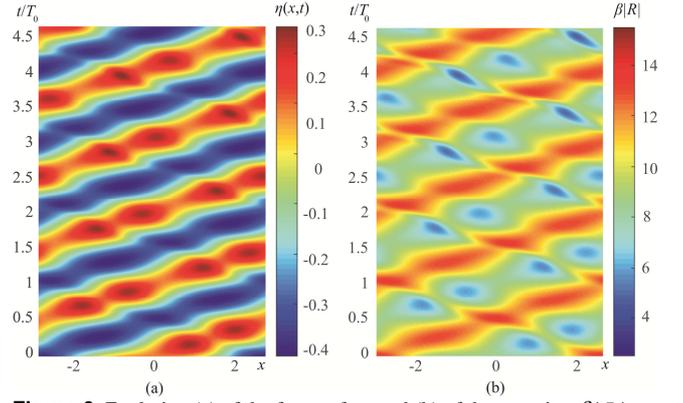

**Figure 5.** (a) The spectrum of the function $Y(u,t)$ is shown at successive instants of time $t_1=0$, $t_2/T_0=2.35$, $t_3/T_0=4.48$, the solid line corresponds to the power-law function $|Y_k| \sim k^{-2}$. (b) The maximum of the surface curvature versus time.

Thus, the numerical simulation of the interaction between counter-propagating nonlinear waves in the framework of the reduced system (8) demonstrates a tendency to the formation of singularities at the fluid surface; the regions appear where the pressure exerted by electric undergoes a discontinuity and the dynamic pressure increases almost an order of magnitude. It is interesting to note that the interaction of the contrary-law-propagating Alfven waves leads to the acceleration of plasma particles [28].

It should be noted that the observed behavior of the fluid surface sufficiently differs from the simple wave breaking, which is shown in figure 2. For the breaking process, infinite gradients of pressure in the $x$-projection arise only on the geometric reasons even in the case of their continuous distributions along the boundary. The evidence of the singular behavior is the spectral functions of $Y$: they tend to the power-law distribution, $|Y_k| \sim k^{-2}$. A possible reason of the formation of these singularities is the absence of terms responsible for capillarity and gravity in the equations (8). Further, we will consider the interaction of oppositely-traveling nonlinear waves in the framework of the complete system of the evolution equations (5) and (6).

### 5.3 INTERACTION OF WAVES IN THE CASE OF FINITE FIELD

For numerical simulation of the wave interaction on the basis of the complete equations system (5) and (6), we take the initial condition for the function $R$ in the form (10). The initial condition for $V$ is modified to the following:

$$V(u,0) = -A_1 i V_p(k_1)\exp(-ik_1u) + A_2 i V_p(k_2)\exp(-ik_2u)$$

For a correct comparison of the results with the previous experiment, we choose the calculation parameters $A_{1,2}$, $k_{1,2}$, $\beta$ to be the same as earlier. Figure 6 shows the evolution of the liquid surface and local electric field at the boundary; the time period is chosen as $T_0=2\pi/V_p(k_2,\beta)$. One can immediately see the main difference in dynamics of the system from the case of a strong field: the tendency to the formation of a discontinuity in the electric field strength is not observed now. The time interval in figure 6 was chosen to provide convenience of comparison with the figure 4. In fact, the calculation interval has reached a rather large value $t/T_0 \approx 38.43$.

**Figure 6.** Evolution (a) of the free surface and (b) of the quantity $\beta|R|$.

One can see from the figure 6 that the region of the discontinuity formation is smoothed. The reason is that the capillary waves are excited in the region of shock formation. At the developed stages of the system evolution, this process leads to the appearance of a massive cascade of small-scale waves. It can be seen from the figure 7 that, at large times, the dynamics of liquid fundamentally differs from the case of a strong field shown in the figure 3. The capillary waves generation results in the fact that dependencies plotted in the figure look very chaotic. Apparently, the picture observed in the figure 7 evidences that the external tangential electric field can lead to the development and acceleration of the so-called capillary wave turbulence [29].

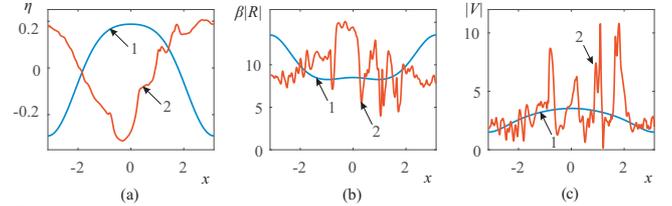

**Figure 7.** (a) The surface of liquid, (b) the local electric field, and (c) the velocity of the fluid surface are shown at the initial moment (blue curves "1") and at the end of calculation interval $t/T_0 \approx 38.43$ (red curves "2").

Figure 8 shows the time dependence of the curvature maximum and spectra of the function $Y(u,t)$ at successive instants of time. At the initial stages of the system evolution, $t/T_0 < 5$, it is possible to distinguish discrete jumps of the curvature as in the figure 5b. At large times, the dependence shown in the figure 8a becomes very complicated. The transition of energy to small scales leads to the error accumulation and, as a consequence, to a bounded computation interval. Indeed, the spectrum of the function $Y(u,t)$ ceases to be localized with time. From the physical point of view, the divergence of the algorithm is related to the absence of viscous forces in the problem under consideration. For the full description of the turbulent motion of the surface, one should introduce additional terms into the equations (5) and (6), as it was done, for instance, in [30]. We plan to consider the turbulent motion of gravity-capillary waves in a tangential electric field and also take into account the terms responsible for the turbulent viscosity in future.

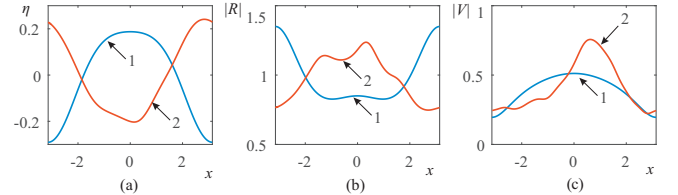

**Figure 10.** (a) The surface of liquid, (b) the quantity $|R|$, and (c) the velocity of the fluid surface are shown at the initial moment (blue curves "1") and at the end of calculation interval $t/T_0 \approx 1.12 \times 10^3$ (red curves "2").

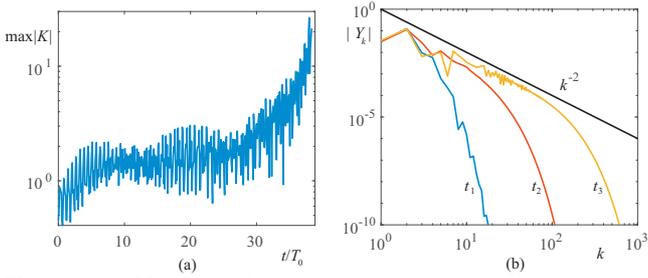

**Figure 8.** (a) Maximum value of the surface curvature versus time, (b) the spectrum of the function $Y(u,t)$ is shown at successive instants of time, $t_1$=0, $t_2/T_0$=2.35, $t_3/T_0$=4.48. The solid line corresponds to the power-law function $|Y_k| \sim k^{-2}$.

In conclusion, let us give an example of the simulation of the interaction between counter-propagating periodic waves in the case, where the external electric field is zero, $\beta$=0. It turns out that in this situation, the generation of capillary waves becomes very slow (it is almost absent in comparison with the previous calculation). Figure 9 shows the evolution of the free surface and of the quantity $|R|$, which does not have a sense of the electric field strength in the latter case. From the figure 9b, one can see that the interaction of waves is not elastic. The wave with large amplitude moves in the positive direction of $x$ axis absorbing the counter-propagating wave with small amplitude. If the electric field is absent, the dynamics of liquid is very stable, the computation interval reached the value, $t/T_0 \approx 1.12 \times 10^3$. At the end of calculation interval, the relative error has not exceeded the small quantity $10^{-8}$. Figure 10 shows the dependencies analogous to that shown in the figures 3 and 7. It can be seen that even after a long interval of time, the plotted functions remain smooth and do not have noticeable small-scale perturbations. It is noteworthy that, during the evolution of the system, a critical increase in the curvature of the surface was not observed, and the spectrum of the function $Y(u,t)$ remained localized. Thus, the mechanism of the formation of massive cascade of small-scale capillary waves, which was observed in the previous numerical experiment, is exactly related to the interaction of counter-propagating nonlinear waves in the presence of a strong tangential electric field.

# 6 CONCLUSION

At the present work, the processes of propagation and interaction of nonlinear waves on the free surface of a non-conducting liquid under the action of gravity, capillarity, and tangential electric field were numerically simulated. The computational algorithm was based on the time-dependent conformal transformation of the region occupied by the fluid into a half-plane. This approach allows to reduce the original spatially two-dimensional problem to the one-dimensional one without loss of generality that essentially increases the efficiency of calculations.

The results of our numerical simulations show that, for sufficiently high electric field strength, the surface waves can separately propagate along, or against the field direction almost without distortions. This result evidences that the limit of a strong field [14-16], in which the gravitational and capillary forces are neglected, indeed can be realized for the waves traveling in one direction (at least for a finite time). Numerical simulations show that in the limit of a strong electric field, the elastic interaction of counter-propagating waves leads to the formation of singularities, viz. points on the liquid surface where the electrostatic and dynamic pressures undergo a discontinuity. The spectrum of the surface tends to the power-law distribution, $|Y_k| \sim k^{-2}$. This indicates to an infinite increase in the curvature of the boundary.

In the case of a finite electric field, the gravitational and capillary forces lead to smoothing the region of pressure discontinuity at the initial stages of the wave interaction. At the developed stages of the system evolution, the small-scale capillary waves are intensively radiated from of the shock front region that causes the chaotic behavior of the system. It is shown that, in the absence of an external field, the wave radiation is very weak; it is not almost noticeable at the comparable time scales. Thus, the investigated mechanism of the counter-propagating waves interaction resulting in the formation of singularities in the strong field limit can enhance and accelerate the development of capillary turbulence of the liquid surface.

## ACKNOWLEDGMENT

This work was supported by the Ministry of Education and Science of the Russian Federation (state contract No. 0389-2014-0006). The work of E.A.K. was supported jointly by RFBR project No. 16-38-60002_mol_a_dk and by the

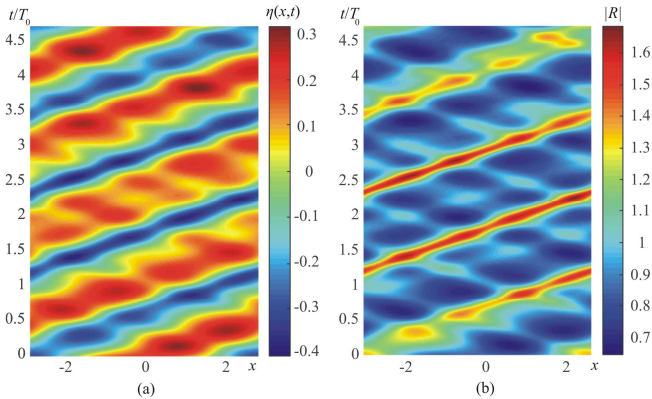

**Figure 9.** The evolution (a) of the free surface and (b) of the quantity $|R|$.

Presidential Programs of Grants in Science (project No. SP-32.2016.1). The work of N.M.Z. was supported by the RFBR (project Nos. 16-08-00228, 17-08-00430) and by the Presidium of UB, RAS (project Nos. 15-8-2-8, 18-2-2-15).

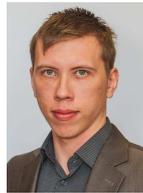

**Evgeny A. Kochurin** was born in 1988. He received M.Sc. degree in electrical physics from Ural Federal University, in 2011. Since 2010 he has been employed at the Institute of Electrophysics, Ural Branch of Russian Academy of Sciences, Yekaterinburg city. He received Candidate of Science degree (PhD) form the Institute, in June, 2015. His PhD thesis was devoted to theoretical investigation of nonlinear dynamics of free and contact boundaries of dielectric liquids under the action of strong electric field.

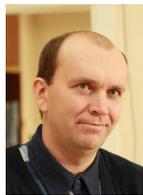

**Nikolay M. Zubarev** was born in 1971. He received the M.Sc. degree in applied mathematics and physics from Moscow Institute of Physics and Technology, Moscow, Russia, in 1994, the Cand.Sci. degree from the Institute of High Current Electronics, Russian Academy of Sciences, Tomsk, Russia, in 1997, and Dr.Sci. degree from the Institute of Electrophysics, Russian Academy of Sciences, Ekaterinburg, Russia, in 2003. He is currently with the Institute of Electrophysics, Russian Academy of Sciences, where he is involved in theoretical studying of nonlinear phenomena in liquids with free surface under the action of an electric field and electrical discharges in gas and vacuum. Prof. Zubarev is a Corresponding Member of the Russian Academy of Sciences since 2016. He was a recipient of the State Prize of the Russian Federation in Science in 2003.